\documentclass{aastex62}
\usepackage{amssymb}
\usepackage{amsmath} \usepackage{lipsum}
\usepackage{amsmath,amsthm}
\graphicspath{{./}{figures/}}

\received{February 22, 2019}
\revised{March 25, 2019}
\accepted{March 25,2019}
\submitjournal{ApJ} 

\shorttitle{Investigation of RBV2}
\shortauthors{Seligman and Shariff}

\begin{document}

\title{Investigation of a Vorticity-preserving Scheme for the Euler Equations}

\correspondingauthor{Darryl Seligman}
\email{darryl.seligman@yale.edu}

\author{Darryl Seligman}
\affil{Astronomy Department, Yale University,
    New Haven, CT 06511, USA}

\author{Karim Shariff}
\affiliation{NASA Ames Research Center
Moffett Field, CA 94035, USA}



\begin{abstract}
We investigate the vorticity-preserving properties of the compressible, second-order residual-based scheme, ``RBV2''. The scheme has been extensively tested on hydrodynamical problems, and has been shown to exhibit remarkably accurate results on the propagation of inviscid compressible vortices, airfoil-vortex interactions on a curvilinear mesh, vortex mergers in an astrophysical accretion disk, and the establishment of a two-dimensional inverse cascade in high-resolution turbulent simulations. Here, we demonstrate that RBV2 sustains the analytic solution for a one-dimensional shear flow. We assess the fidelity by which the algorithm maintains a skewed shear flow, and present convergence tests to quantify the magnitude of the expected numerical dispersion. We propose an adjustment to the dissipation in the algorithm that retains the vorticity-preserving qualities, and accurately incorporates external body forces, and demonstrate that it indefinitely maintains a steady-state hydrostatic equilibrium between a generic acceleration and a density gradient. We present a novel
numerical assessment of vorticity preservation for discrete-wavenumber vortical modes up to the Nyquist wavenumber. We apply this assessment to RBV2 in order to quantify the extent to which the scheme preserves vorticity for the full Euler equations.  We find that RBV2 perfectly preserves vorticity for modes with symmetric wavenumbers, i.e., $k_x=k_y$, and that the error increases with asymmetry. We simulate the dynamical interaction of vortices in a protoplanetary disk to demonstrate the utility of the updated scheme for rendering astrophysical flows replete with vortices and turbulence.  We conclude that RBV2 is a competitive treatment for evolving vorticity-dominated astrophysical flows with minimal dissipation.
\end{abstract}



\section{Introduction} \label{sec:intro}
Simulations of high Reynolds number flows that respect the evolution of vorticity, and minimize its numerical diffusion, are of key importance for many applications of astrophysical fluid dynamics.  Numerical schemes that accurately treat vorticity are of special importance for high-value problems such as blade-vortex interactions in aeronautical engineering. Any numerical investigation of flows with high Reynolds numbers, where turbulence and vortices are expected to dominate the dynamics, inherently require a numerical method that accurately evolves the vorticity, as well as the density, momenta, and energy in the flow. In the astrophysical context, hydrodynamical or magnetohydrodynamical turbulence is thought to be the main driver of the dynamics in convective domains such as Jupiter's atmosphere \citep{Marcus1988}, in accretion disks \citep{Balbus1991} and in protostellar disks \citep{Shariff2009, Lyra2018}, to name a few examples.

The Euler equations, and the vorticity equation derived from them, have no dissipative terms.  However, when solving the Euler equations numerically on a finite mesh, a certain amount of dissipation is required to represent the transfer of kinetic energy and enstrophy from  resolved to  unresolved scales.
On the other hand, numerical simulations generally suffer from excessive energy and enstrophy dissipation such that their spectra begin to rapidly fall-off at length scales substantially larger than the numerical grid size.
This is due to numerical dissipation, which is either inherent in the method (such as upwinding), or explicitly added to guarantee a stable solution. Schemes used to capture shock-waves employ
nonlinear dissipation. Particle-based schemes also tend to diffuse vorticity due to the dissipation associated with the adopted smoothing kernel. A classical approach to decrease diffusivity, and thereby increase the effective Reynolds number in the flow solution, is to increase the resolution of the simulation. Since hydrodynamical simulations are computationally expensive, it is of interest to investigate alternative approaches to controlling unphysical vorticity diffusion. 

Theoretically, since enstrophy cascades to small scales, long  simulations  must allow for unresolved vorticity structures. Therefore, simulations where turbulence and vortices are expected to dominate the dynamics should aim to model the effect of the unresolved scales on the resolved scales. The first model of this kind was proposed by  \cite{Smagorinsky1963}, and from comparisons with direct numerical simulations it is known that it is too dissipative. Recently there has been an effort to develop less dissipative models \citep[][and references therein]{Rozema2015}. Models  should ideally dissipate enstrophy  only at the scales near the grid cut-off where non-linear interactions between the resolved and unresolved motions are important. 

\citet{Morton2001} introduced the concept of vorticity-preservation. A scheme is said to be vorticity-preserving if the evolution equation for discrete vorticity looks the same as the continuous vorticity equation discretized by central difference operators. This requires that the numerical operators, denoted by a tilde, for curl and gradient satisfy $(\widetilde{\mathrm{curl}}) (\widetilde{\mathrm{ grad}}) = 0$, which holds if the derivative operators for the different spatial directions commute.  It also requires that upwind and/or dissipation terms have zero $\widetilde{\mathrm{curl}}$. They demonstrated that the Lax-Wendroff-Ni scheme \citep{Ni1982}, exactly preserves the discrete vorticity for the Euler equations in the case of linearized acoustics in a uniform medium.

\citet{Lerat2007} presented the numerical scheme RBV2, which extended the concept of vorticity preserving numerical simulations to the linearized (about a uniform flow) and full compressible Euler equations. The method incorporates additional dissipation into a central difference scheme, which by construction, respects the vorticity evolution.  The RBV2 algorithm has been extensively tested on hydrodynamical problems. \citet{Lerat2007} showed that the scheme is able to perfectly sustain the two-dimensional, compressible inviscid vortex proposed by \citet{Yee1999}. \citet{Seligman2017} compared the same test problem with commonly used astrophysical codes, and found that RBV2 remarkably outperformed a Godunov-type solver using third-order Runge–Kutta timestepping and the weighted essentially nonoscillatory WENO3 \citep{Liu1994} reconstruction scheme, a van-Leer advection based code and a spectral code. \citet{Falissard2008} demonstrated that the RBV2 scheme performs favorably on a linearly advected vortex and demonstrated its fidelity on the computation of airfoil-vortex interactions on a curvilinear mesh. \citet{Seligman2017} also demonstrated that, in the comparison codes, RBV2 was the only code among those tested that was capable of establishing and sustaining an inverse cascade in an unforced two-dimensional turbulence simulation. Motivated by the extremely efficient test results for RBV2, \citet{Seligman2017} implemented the RBV2 scheme in the context of astrophysical accretion disks. They found that it ran at a much higher effective Reynolds number than the Godunov method for comparable computational effort, based on a test of vortex merger with a background shear. They concluded that the algorithm may have important ramifications for simulations of astrophysical environments where turbulence was expected to dominate the dynamics, and where strong shocks are unlikely to be present.

It is of interest to note that there are other methods to solve the Euler equations that are capable of running at  high Reynolds numbers, and which minimze the diffusion of vorticity. \citet{Seligman2017} found that the spectral code Dedalus \citep{Burns2016} was able to  sustain the inviscid vortex, although it took ten times longer to run than the RBV2 algorithm as a consequence of the very small time step required to run at very high Reynolds number.  

\citet{Falissard2008}, however, noted that when RBV2 is used to model flows that contain significant advection diagonally through the grid, spurious short-wavelength perturbations may manifest in
the crosswise direction. In order to dissipate these perturbations, while maintaining the desirable vorticity-preserving quality of RBV2, \citet{Seligman2017} implemented high-order spatial filters as proposed by \citet{Lele1992}. The details of this implementation may be found in Section 2 of \cite{Seligman2017}.

In this paper, we continue to explore the potential of the RBV2 scheme, and specifically, we conduct an investigation of the vorticity-preserving qualities of the algorithm. We present several variations of a shear flow as a  hydrodynamical test, all ran to the same simulation time. These tests provide a linear wave convergence test, and quantify the magnitude of the numerical dispersion in the method. We propose an update to the algorithm to incorporate external forces, which is necessary for most astrophysical applications, such as numerical simulations of a protoplanetary disk or a planetary atmosphere. We implement this update, and demonstrate that the new algorithm is able to sustain a shear flow in the presence of a vertical hydrostatic equilibrium. We present  a novel assessment of vorticity preservation, which we apply to the algorithm. We conclude with  a simulation of the dynamical interaction of seven vortices in a protoplanetary disk, to demonstrate the utility of the updated scheme for rendering turbulent astrophysical flows. 

The plan of this paper is as follows. In $\S$ 2, we present an overview of the algorithm, extending the formal analysis of its merits (and shortcomings), and proceeding from the work of \cite{Lerat2007}. In $\S$ 3 we briefly present the numerical methods used to implement the scheme. In $\S$ 4 we present the numerical and analytical results for RBV2 on a one-dimensional shear flow.  In $\S$ 5, we present the results of numerical experiments on a skewed shear flow. In $\S$ 6, we propose an update to the algorithm to incorporate external body forces and demonstrate that the updated scheme sustains a hydrostatic equilibrium. In $\S$ 7, we provide a numerical quantification for the dissipation of vorticity for a set of vortical modes with wavenumbers varying from 1 to the Nyquist wavenumber. In $\S 8$, we perform shearing sheet simulations of vortex interactions in a protoplanetary disk as a demonstration of the updated scheme. In $\S 9$, we discuss the ramifications of these results, outline future prospects, and conclude.  

\section{Overview of the Algorithm}

Following the nomenclature of \citet{Lerat2007}, we write the two-dimensional Euler equations as,

\begin{equation}\label{generalsystem}
    \partial_t {\boldsymbol w} +\partial_x {\boldsymbol f}+\partial_y {\boldsymbol g}=0 \, , 
\end{equation}
for three general vectors ${\boldsymbol w}$, ${\boldsymbol f}$ and ${\boldsymbol g}$. The exact form of these three vectors depends on the equation set to be solved; \citet{Lerat2007} consider three sets of vectors for the cases of pure acoustics, the linearized Euler equations, and the full Euler equations. Since they will be used frequently in this paper, we provide the latter two here. For the Euler equations linearized around a uniform flow $(u_0,v_0)$, the vectors are defined as:
\begin{equation}\label{lineareuler}
\begin{aligned}
    {\boldsymbol w} &= \begin{bmatrix}
           p \\
            u \\
            v
         \end{bmatrix}\, ,
    {\boldsymbol f} &= \begin{bmatrix}
           pu_0+u \\
           uu_0+p \\
           vu_0
         \end{bmatrix}\, ,
         {\boldsymbol g} &= \begin{bmatrix}
           pv_0+v \\
           uv_0 \\
            vv_0+p
         \end{bmatrix}\, ,
\end{aligned}
\end{equation}
whereas for the full Euler equations, they are given by, 
\begin{equation}\label{fulleuler}
\begin{aligned}
    {\boldsymbol w} &= \begin{bmatrix}
           \rho \\
           \rho u \\
           \rho v \\
           \rho E
         \end{bmatrix}\, ,
    {\boldsymbol f} &= \begin{bmatrix}
           \rho u \\
           \rho u^2 + p \\
           \rho uv \\
           (\rho E + p)u
         \end{bmatrix}\, ,
         {\boldsymbol g} &= \begin{bmatrix}
           \rho v \\
           \rho uv \\
           \rho v^2 +p \\
           (\rho E +p)v
         \end{bmatrix}\, .
\end{aligned}
\end{equation}
For simplicity, we adopt the polytropic equation of state, $p=c_s^2\rho^\gamma$. For the remainder of this paper, we will further specialize to an isothermal equation of state, i.e., $\gamma=1$, with a uniform sound speed.

The RBV2 scheme may be cast in the form, 
\begin{equation}\label{LeratScheme}
  D_0 \mu_0 {\boldsymbol w}+ D_1 \mu {\boldsymbol f}+ D_2 \mu {\boldsymbol g}=h_1 D_1 (\Phi_1 \tilde{{\boldsymbol q }})+h_2 D_2 (\Phi_2 \tilde{{\boldsymbol q}}) \, ,
\end{equation}
where $\mu_0 = \mu^2$ is a grid-averaging operator and $\mu$ will be defined below.
The terms on the right-hand-side of (\ref{LeratScheme}) represent the added dissipation, $h_1,h_2$ being positive parameters tending towards zero with increased resolution. $\Phi_1$ and $\Phi_2$ are matrices with the same eigenvectors as the Jacobians $A=df/dw$ and $B=dg/dw$ respectively. The general difference operators are $D_0$ , $D_1$, and $D_2$, which represent the time, $x$ and $y$ numerical derivatives respectively, and the general averaging operators are $\mu$ and $\mu_0$. $\tilde{{\boldsymbol q}}$ is, in the general case, a vector involving discrete functions of ${\boldsymbol w}$, ${\boldsymbol f}$ and ${\boldsymbol g}$, although for the RBV2 algorithm, $\tilde{{\boldsymbol q}}=D_0\mu {\boldsymbol w}+D_1{\boldsymbol f}+D_2 {\boldsymbol g}$. For the remainder of the paper,  $\mu=\mu_1\mu_2$, $D_0=\frac{1}{\Delta t}\bar{\Delta}$, $D_1=\frac{1}{\delta x}\delta_1\mu_2$ and $D_2=\frac{1}{\delta y}\delta_2\mu_1$. These operators are defined on a Cartesian mesh with $x_j=j\delta x$,  $y_k=k\delta y$,  with time steps $t^n=n\Delta t$, for an arbitrary primitive variable $\psi$, as,

\begin{equation}\label{operators}
\begin{aligned}
    (\delta_1 \psi)_{j+\frac12,k}&=\psi_{j+1,k}-\psi_{j,k}\\
    (\delta_2 \psi)_{j,k+\frac12}&=\psi_{j,k+1}-\psi_{j,k} \\
    (\mu_1 \psi)_{j+\frac12,k}&=\frac12(\psi_{j+1,k}+\psi_{j,k}) \\
    (\mu_2 \psi)_{j,k+\frac12}&=\frac12(\psi_{j,k+1}+\psi_{j,k}) \\ 
    (\bar{\Delta} \psi)^{n+1}&=\frac32 \psi^{n+1} - 2\psi^{n} +\frac12 \psi^{n-1} \, .
\end{aligned}
\end{equation}

\citet{Lerat2007} primarily consider $\Vec{\Omega}\equiv\nabla\times\rho\Vec{u}$, as the quantity of interest and refer to it as the vorticity.  By vorticity-preservation they mean that the evolution equation for its discrete analog be identical to the equation obtained when the same difference operators are applied to the continuous evolution equation for $\Vec{\Omega}$.  We adopt the same terminology and refer to $\vec{\omega} \equiv\nabla\times\Vec{u}$ as the proper vorticity.

The discrete vorticity evolution equation  comes from taking the curl of the momentum part of Equation \ref{LeratScheme},
\begin{equation}\label{vorticiy_evolution}
\begin{split}
D_0\mu_0 \widetilde{\mathrm{curl}}_w+D_1\mu \widetilde{\mathrm{curl}}_f+D_2 \mu \widetilde{\mathrm{curl}}_g\\ = h_1D_1\widetilde{\mathrm{curl}}_1+h_2D_2\widetilde{\mathrm{curl}}_2\, ,
\end{split}
\end{equation}
where
\begin{equation}\label{curldefinitions}
\begin{aligned}
    \widetilde{\mathrm{curl}}_w&=D_1 {\boldsymbol w}^{(3)} -D_2 {\boldsymbol w}^{(2)} \\
    \widetilde{\mathrm{curl}}_f&=D_1 {\boldsymbol f}^{(3)} -D_2 {\boldsymbol f}^{(2)} \\
    \widetilde{\mathrm{curl}}_g&=D_1 {\boldsymbol g}^{(3)} -D_2 {\boldsymbol g}^{(2)} \\
    \widetilde{\mathrm{curl}}_m&=D_1(\Phi_m \tilde{{\boldsymbol q}})^{(3)}-D_2(\Phi_m \tilde{{\boldsymbol q}})^{(2)}\,,\,m = 1, 2 .
\end{aligned}
\end{equation}

Theorems 1--5 in \citet{Lerat2007} prove that the scheme given by Equation \ref{LeratScheme}, with the condition,
\begin{equation}\label{Condition15}
\begin{aligned}
    h_1\Phi_1&=h A \\ 
    h_2 \Phi_2&=hB \, ,
\end{aligned}
\end{equation}
for some constant $h$, is vorticity preserving for the case of pure acoustics and the  Euler equations linearized around a uniform advecting flow. Without repeating the details of each proof, we note that in general, they adhere to the following structure: (i) Define the scheme, (ii) Solve for the vorticity evolution (Equation \ref{vorticiy_evolution}) within the assumptions and (iii) Show that the all terms in the right hand side (RHS henceforth) of Equation \ref{vorticiy_evolution} reduces to zero. The extra dissipation here, by construction, demands that the discrete solution to the Euler equations respects the  vorticity equation to machine precision.


Theorem 6 in \citet{Lerat2007} states that the Scheme \ref{LeratScheme} with Condition \ref{Condition15}, preserves vorticity for the linearized Euler Equations (Equation \ref{lineareuler}), if the discrete operator,

\begin{equation}\label{lamdba}
    \Lambda = \mu - h_1D_1\Phi_1-h_2D_2\Phi_2,
\end{equation}
is non-singular. For the linearized system, the discrete vorticity evolution, Equation \ref{vorticiy_evolution}, reduces to 

\begin{equation}
(D_0\mu_0+u_0D_1\mu+v_0D_2 \mu) \tilde{\omega}=h_1D_1\widetilde{\mathrm{curl}}_1+h_2D_2\widetilde{\mathrm{curl}}_2\, ,
\end{equation}
where the numerical (proper) vorticity $\tilde\omega \equiv D_1v-D_2u$.
They show that the Scheme \ref{LeratScheme} is equivalent to the matrix equation,

\begin{equation}
    \Lambda \tilde{{\boldsymbol q}} = 0 \, .
\end{equation}
Therefore, if $\Lambda$ is invertible, this yields $\tilde{{\boldsymbol q}}=0$. They conclude that, since  the RHS of Equation \ref{vorticiy_evolution} is then zero,  the scheme is vorticity-preserving. However, we note that if $\tilde{{\boldsymbol q}}=0$,  the RHS of the original scheme (\ref{LeratScheme}) is identically zero; in other words, the additional dissipation in the original scheme vanishes.  This raises two questions: (i) why  is it necessary to introduce the additional dissipation in the first place, and (ii) would its absence mean that shocks would not be captured without spurious oscillations?
%
Theorem 7 in \citet{Lerat2007} extends Theorem 6 to the case of the full Euler equations, and leads to the same issue of the added dissipation being zero.
It was this issue that led us to numerically investigate the algorithm; see \S\ref{sec:assessment}.
There we demonstrate that RBV2 does not preserve vorticity in general, which implies that the added dissipation terms are non-zero in general.

However, there is now a large amount of numerical evidence that the RBV2 algorithm is considerably less dissipative than many schemes.  We therefore conducted the following investigations to elucidate the extent to which the scheme preserves vorticity. 

\section{Solution of the Implicit System}
In this section, we describe the numerical method by which we solve the implicit system (\ref{LeratScheme}).  For a more detailed description, we refer the reader to \citet{Seligman2017}.  Equation (\ref{LeratScheme}) is written as a residual equal to zero, i.e., ${\bf R}({\bf w}) = 0$.
We solve this equation by stepping the following equation to steady-state (in fictitious time) using the technique proposed by \citet{Jameson1991}:
\begin{equation}\label{eq:t*}
    \frac{\partial{\bf w}}{\partial t^*} +{\bf R}({\bf w})=0 \, ,
\end{equation}
where $t^*$ is a fictitious time. The residual is split into its convective and dissipative parts (${\bf Q}$ and ${\bf D}$, respectively),

\begin{equation}\label{eq:res}
{\bf R}({\bf w})={\bf Q}({\bf w})+{\bf D}({\bf w})\, .
\end{equation}
The solution vector at subsequent time steps ${\bf w}^{n+1}$ is found by iterating over the following algorithm, where the superscript $k$ symbolizes successive iterations for ${\bf w}$,

\begin{equation}\label{eq:guess1}
{\bf w}^{n+1,0}={\bf w}^n \, ,
\end{equation}
and successive iterations are
\begin{equation}\label{eq:wguess}
{\bf w}^{n+1,k}={\bf w}^n-\alpha_k\Delta t^*[{\bf Q}^{k-1}+{\bf D}^{k-1}]\, .
\end{equation}

The initial convective and dissipative residuals are calculated from the previous time step, ${\bf Q}^0={\bf Q}[{\bf w}^n]$ and ${\bf D}^0={\bf D}[{\bf w}^n]$. The $k$th residuals are calculated using ${\bf Q}^k={\bf Q}[{\bf w}^{n+1,k}]$ and ${\bf D}^k=\beta_k{\bf D}[{\bf w}^{n+1,k}]+(1-\beta_k){\bf D}^{k-1}$. The coefficients $\alpha_k$ and $\beta_k$ are chosen to maximize the stability interval along the imaginary and negative real axis of each Fourier mode of the solution. \citet{Jameson1991} reported  that  effective choices for $\alpha$ and $\beta$ are $\alpha_1=1/3$, $\alpha_2=4/15$,  $\alpha_3=5/9$, $\alpha_4=1$ and $\beta_1=1$, $\beta_2=1/2$, $\beta_3=0$, and $\beta_4=0$. For stability, in all simulations presented in this paper, we use a global timestep that is one third of that given by the Courant condition, e.g. that $\Delta t = \frac13 \Delta x/\sqrt{u^2+v^2+c_s^2}$, where $u$ and $v$ are the maximum velocities at any point in the grid, and we use three sub-iterations of the scheme at each step. 

\begin{figure}
\begin{center}
\includegraphics[angle=0,trim={0.1cm 0.1cm 0.1cm 0cm},scale=.99]{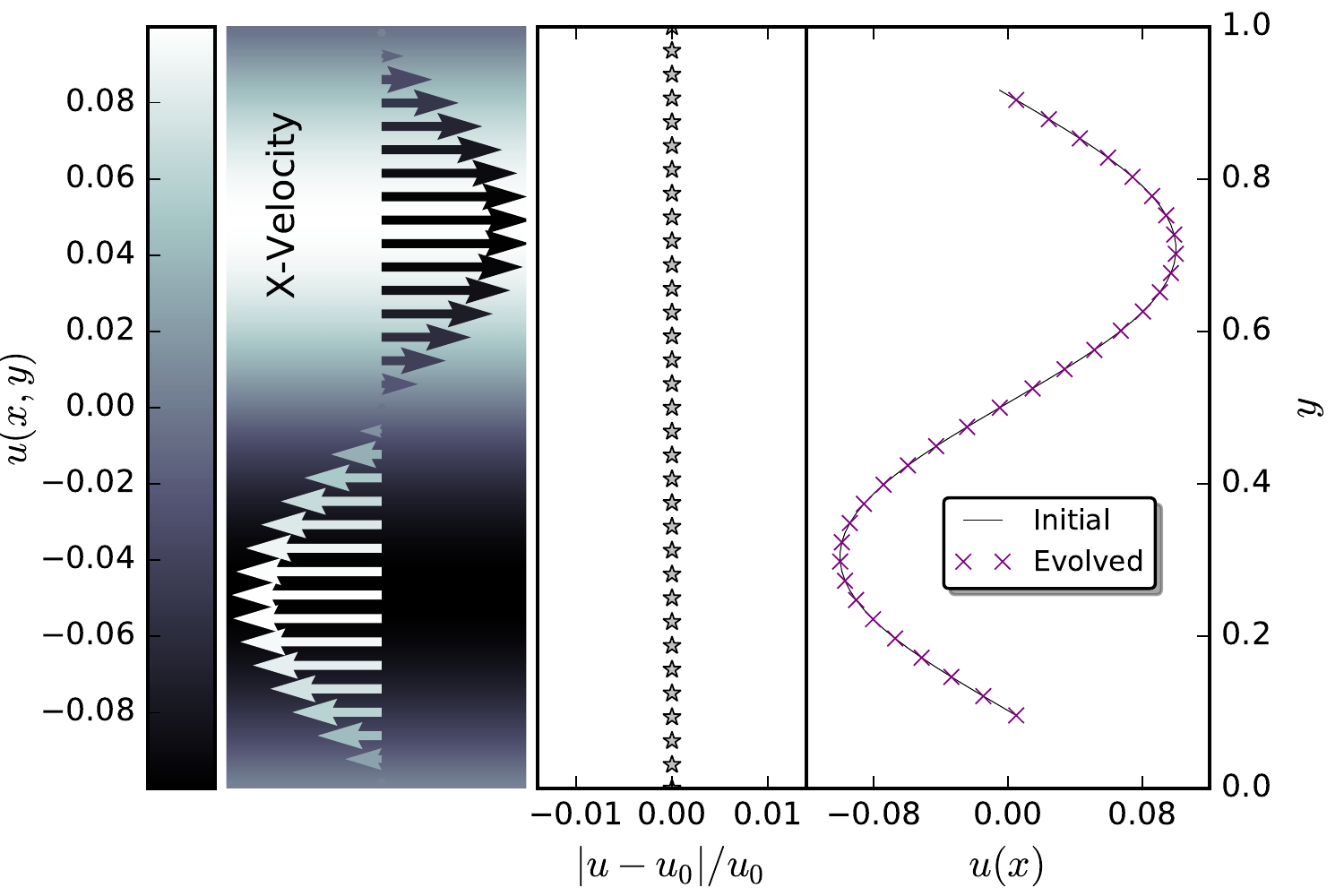}
\figcaption{The fidelity with which RBV2 maintains a one-dimensional shear flow. The initial conditions for the simulations are shown in the left  panel. The color scale in the image corresponds to the $x$-velocity, $u(y)$. The overlaid arrows give a visual representation of the velocity field. In the right panel, the initial, analytic solution is shown in the grey solid line, and the numerical solution evolved for $t\sim60$ sound crossing times is shown in purple $\times$'s. The corresponding residuals, $(u-u_0)/u_0$, are shown in grey stars in the middle panel. We initialize the simulation using Equations \ref{1dshear_IC} on a $64\times64$ zone grid. The amplitude of the shear flow is $U=0.1$, the sound speed is $c_s=1$ and the domain length is $L_x,L_y=1$. It is evident that the scheme is maintaining the analytic solution.}
\label{fig:1dshear}
\end{center}
\end{figure}

\section{One-Dimensional Shear Flow}
In this section we examine both analytically and numerically, the fidelity with which RBV2 applied to the full Euler equations (\ref{fulleuler}) maintains the one-dimensional shear flow:

\begin{equation}\label{1dshear_IC}
\begin{aligned}
    u(x,y)&=U\sin(my) \\ 
    v(x,y)&=0 \\
    \rho(x,y)&=\rho\\
    p(x,y)&=c_s^2\rho \, .
\end{aligned}
\end{equation}
Since the equation of state is isothermal, we may neglect the internal energy equation. The resulting Jacobian matrices are,

\begin{equation}\label{Jacobians_isothermal}
\begin{aligned}
    A= \begin{bmatrix}
           u & 1 & 0 \\
           u^2+c_s^2 & 2u & 0  \\
           uv & v & u
         \end{bmatrix}\, ,\,
     B= \begin{bmatrix}
           v & 0 & 1 \\
           uv & v & u  \\
           v^2+c_s^2 & 0 & 2v
         \end{bmatrix}\, .
\end{aligned}
\end{equation}

We demonstrate that the dissipation term in the scheme given by Equation \ref{LeratScheme}, and therefore its discrete curl, is identically zero for the one-dimensional shear flow.  The only nonzero part of $\tilde{q}$ involves the $y$ derivative, namely, $\tilde{q} = D_2g$. Therefore, $(\Phi_2 \tilde{q})^{(2)} = uv D_2(\rho v )+ vD_2(\rho uv ) + u D_2(\rho v^2 +p)$. Each term to the right of this equation reduces to zero for the given initial conditions. Since the right-hand-side of scheme \ref{LeratScheme} is just a derivative of this, the scheme should sustain the initial condition.

In order to evaluate the one-dimensional shear flow numerically, we use Equations \ref{1dshear_IC} as initial conditions an a $64\times 64$ zone grid. We use an amplitude of the shear flow $U=0.1$, sound speed  $c_s=1$ on a domain of $L_x,L_y=1$. We run the simulation until $t\sim60$, which corresponds to 60 sound crossing times. The results are shown in Figure \ref{fig:1dshear}, and it is evident that the scheme is maintaining the analytic solution.

\begin{figure}
\begin{center}
\includegraphics[angle=0,scale=.99]{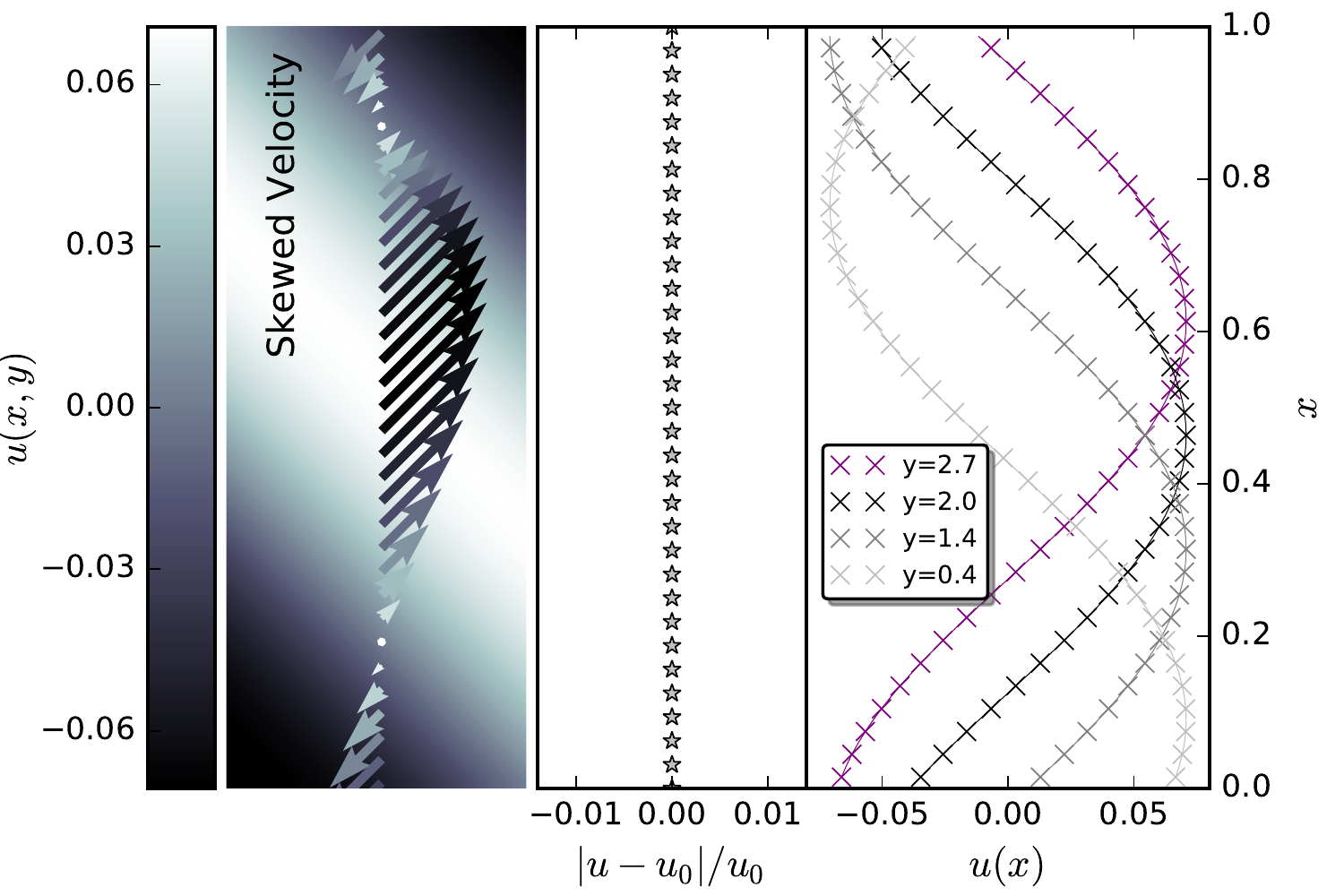}
\caption{The fidelity with which RBV2 maintains a skewed shear flow on a square grid. The initial conditions for the simulations are shown in the left  panel. The color scale in the image corresponds to the $x$ velocity, $u(y)$. The overlaid arrows give a visual representation of the velocity field. In the right panel, the initial analytic solution is shown in the solid lines, and the  numerical solution evolved for $t\sim60$ sound crossing times is shown in $\times$'s.  The different colors represent different $y$ cross sections of the grid. The corresponding residuals, $(u-u_0)/u_0$, are shown in grey stars in the middle panel. We initialize the simulation using Equations \ref{skewed_shear_IC} on a $64\times 64$ zone grid. The amplitude of the shear flow is $A=0.1$, the sound speed  is $c_s=1.0$ and the domain length is $L_x,L_y=\sqrt{2}\pi$. It is evident that the scheme is maintaining the analytic solution. }
\label{fig:skew_shear_evolution}
\end{center}
\end{figure}

\section{Skewed Shear Flow}
\begin{figure}
\begin{center}
\includegraphics[angle=0,scale=.99]{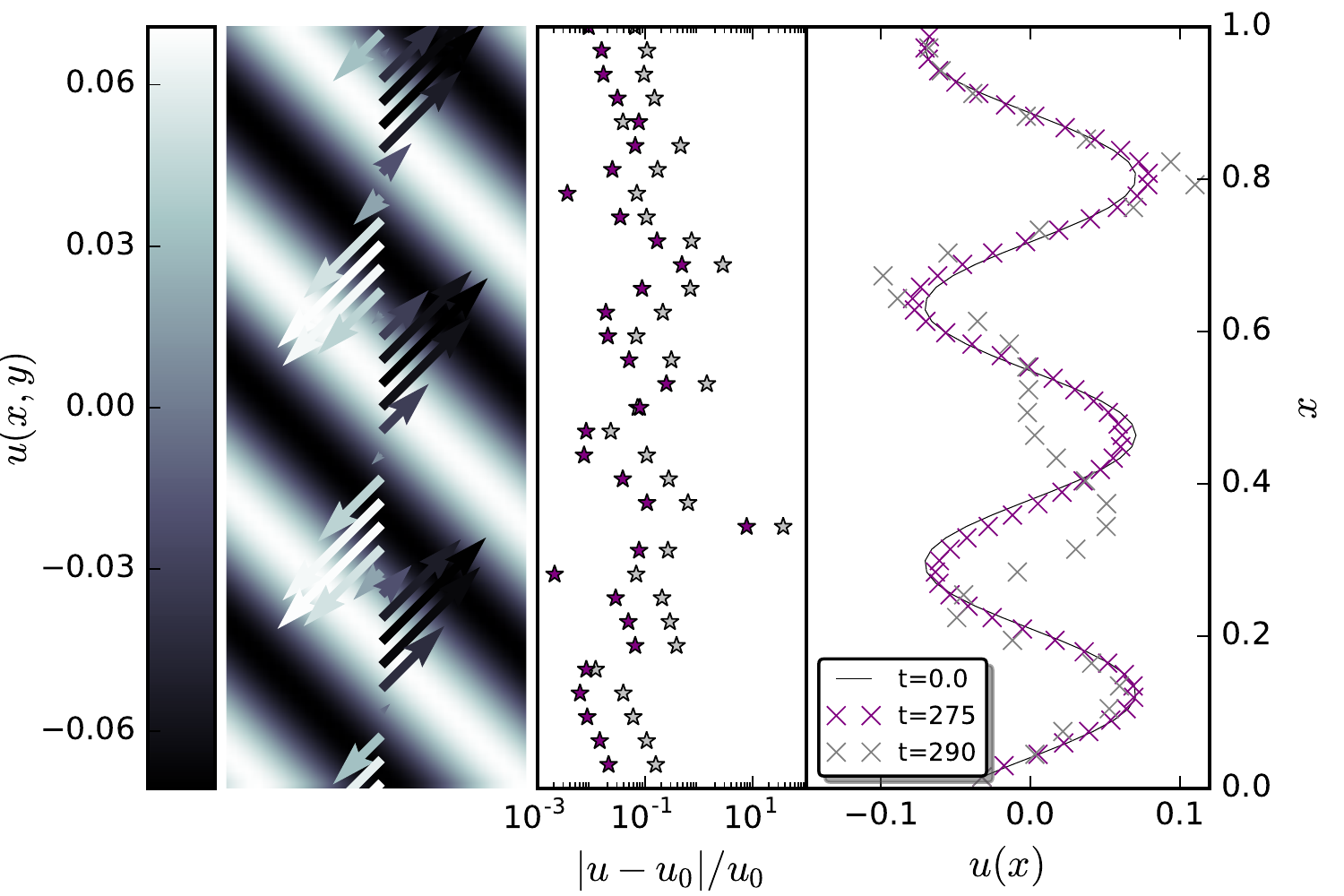}
\caption{The same skewed shear flow experiment as in Figure \ref{fig:skew_shear_evolution}, with a 4$\times$ higher wavenumber. The initial conditions for the simulations are shown in the left  panel. The color scale in the image corresponds to the $x$-velocity, $u(y)$. The overlaid arrows give a visual representation of the velocity field. In the right panel, the initial, analytic solution is shown in the solid lines, and the  numerical solution evolved for $t\sim60$ sound crossing times is shown in dots. The purple $\times$'s show the simulation as the numerical dispersion begins to manifest, and the grey $\times$'s show the solution as the wave begins to display unsteady behavior. The corresponding residuals, $(u-u_0)/u_0$, are shown in grey and purple stars in the middle panel. We initialize the simulation using Equations \ref{skewed_shear_IC} on a $64\times64$ zone grid with four times the wavenumber as in Figure \ref{fig:skew_shear_evolution}. The amplitude of the shear flow is $A=0.1$, the sound speed  is $c_s=1$ and the domain length is $L_x,L_y=\sqrt{2}\pi$. It is evident that the increase in wavenumber increases the numerical dispersion, as expected.}
\label{fig:higher_k_dispersion}
\end{center}
\end{figure}
In this section, we investigate how the RBV2 scheme maintains a skewed shear flow, and present convergence tests to quantify the expected numerical dispersion in the algorithm.  We consider a flow velocity $(u',v')$ in a  coordinate system $(x',y')$, rotated by some angle, $\theta$, with respect to the unprimed coordinates.  The shear flow in the rotated system is  $u'=A\Re\{e^{iky'}\}$. Therefore, in the unrotated (or computational grid) coordinates,

\begin{equation}\label{skewed_shear_IC}
\begin{aligned}
    u(x,y)&=A\cos(\theta)\Re\{ e^{ik(y\cos\theta-x\sin\theta)}\}\\
    v(x,y)&=A\sin(\theta) \Re\{e^{ik(y\cos\theta-x\sin\theta)} \}\\
        \rho(x,y)&=\rho\\
    p(x,y)&=c_s^2\rho \, .
\end{aligned}
\end{equation}
We evolve these initial conditions on a $64\times 64$ grid, with $L_x=L_y=\sqrt{2}\pi$, and run the simulation for ~60 sound crossing times, to be consistent with the one-dimensional shear flow analysis. We skew the shear flow by an angle $\theta=\pi/4$. The results are shown in Figure \ref{fig:skew_shear_evolution}.

As shown in the residuals in Figure \ref{fig:skew_shear_evolution}, there is  no dispersion in the skewed shear flow after 60 sound crossing times to machine precision. Assuming that the discrete vorticity equation has no dissipative terms,  if central difference schemes are used, then the numerical discretization does not produce
any dissipation of vorticity. In other words, only even-order derivatives are present when a truncation error analysis of the difference operators are made, or, one has a purely imaginary modified wavenumber.  However, the difference
operators do have dispersion error because their truncation error has odd-order derivatives and the
modified wavenumber for the first derivative deviates from being linear. This means that when compared to the analytic solution, there should be dispersion, and that the higher wavenumbers experience more rapid dispersion. 

In order to quantify the magnitude of this  numerical dispersion, we present a simulation of the same test with a higher wavenumber, which also may be thought of as a lower resolution simulation. In Figure \ref{fig:higher_k_dispersion}, we increase the wavenumber by a factor of 4. We run this simulation also to 60 sound crossing times, and observe that as the dispersion manifests, the solution begins to go unstable, and the residuals grow to a factor of unity.

\section{Updated Scheme for External Body Forces}
\begin{figure}
\begin{center}
\includegraphics[angle=0,scale=.99]{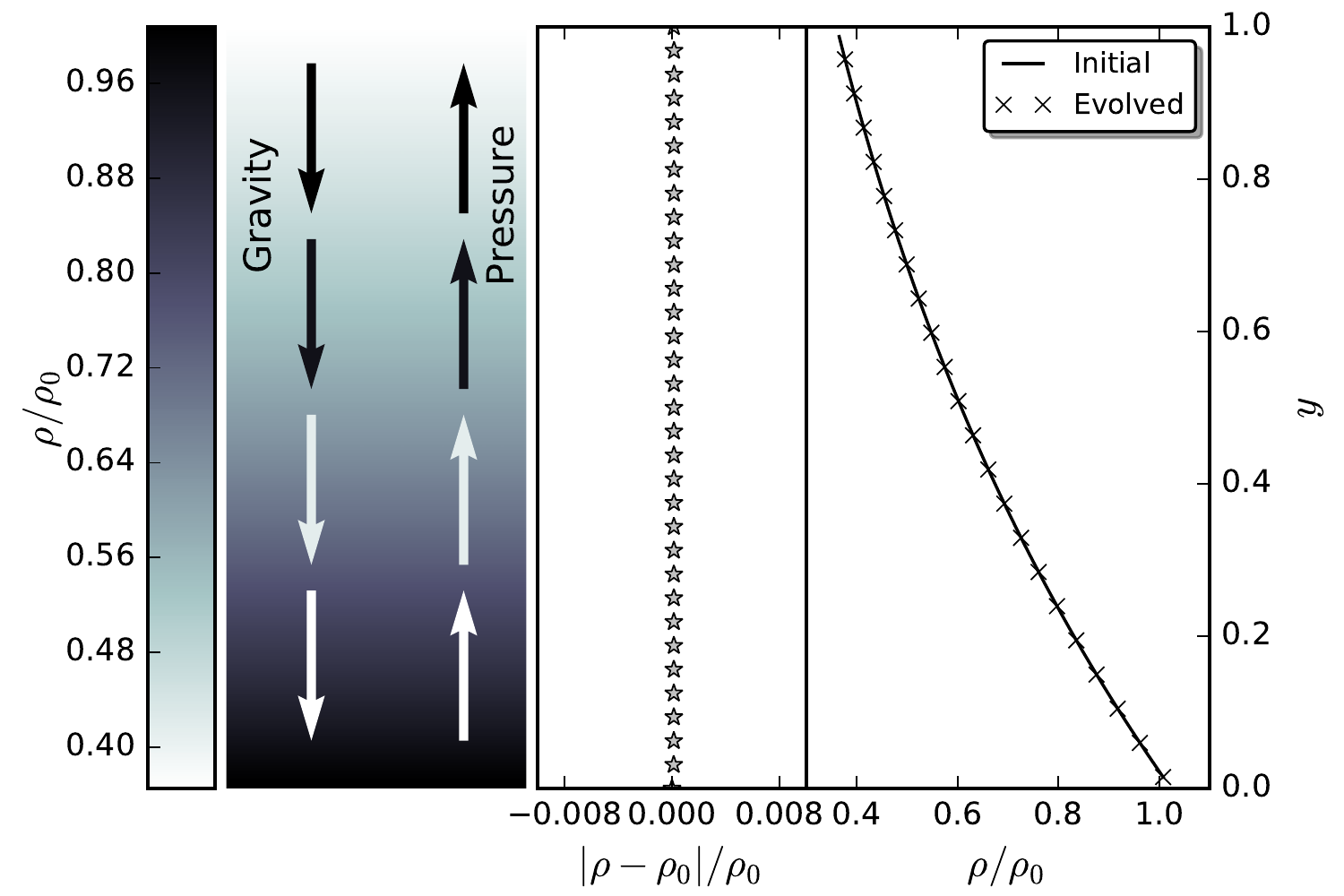}
\caption{The evolution of a shear flow in the presence of a vertical hydrostatic equilibrium. The initial conditions for the simulation, given by Equation \ref{IC_shear_hydrostatic}, are shown in the left  panel. The color scale in the image corresponds to the normalized density, $\rho/\rho_0$. The overlaid arrows give a visual representation of the magnitude and direction of the gravitational body force balanced by the pressure induced by the density gradient. In the right panel, the initial, analytic solution for the density is shown in the solid  line, and the  numerical solution evolved to 50 sound crossing times is shown in $\times$'s. The corresponding residuals, $(\rho-\rho_0)/\rho_0$, are shown in grey  stars in the middle panel. The simulation was run on a 64$\times$64 zone grid, where $L_x=L_y=1$, $c_s=1$, $\rho_0=1$, and  $g=-1$. It is evident that with the proposed update, the scheme is able to maintain the hydrostatic equilibrium imposed by the external body forces.   }
\label{fig:hydrostatic}
\end{center}
\end{figure}
Many astrophysical applications require the advent of external body forces in numerical calculations. For example, a simulation of a protostellar disk requires the incorporation of both a coriolis deflection and an external gravitational field. However,  balancing source terms with non-zero flux gradients is known to be a difficult issue numerically \citep{LeVeque1998}, even in the simplest case where the source terms cancel the flux gradients, such as a vertical hydrostatic equilibrium. We conducted numerical experiments that demonstrated that the RBV2 algorithm, when applied to a naive (cell-centered) source term discretization of an additional, constant gravitational force balanced by a pressure gradient, was unstable. Since the scheme was not developed to incorporate external body forces, it is possible to modify it to handle these and still retain minimal vorticity dissipation.

If a generic external acceleration $a_{ext}=(0,a_x,a_y)$, was added to $\tilde{q}$, e.g.
\begin{equation}
    \tilde{q}=D_0\mu w+D_1 f+D_2 g+\mu\rho  a_{ext} \, ,
\end{equation}
the modified scheme  correctly incorporates the external force.  The external acceleration may be constant or spatially varying, and could possibly represent an external gravitational field. Furthermore, $a_{ext}$ and $\rho$ are defined on grid points, $(j,k)$, so we include the $\mu$ operator so that each term in the residual is evaluated at the midpoints $(j+\frac12,k+\frac12)$.

This new formula for the residual yields a straightforward inclusion of external forces in the scheme that may still be written as $\Lambda \tilde{{\boldsymbol q}} = 0$. The proposed update requires that the difference and average operators commute.


We investigate how the updated scheme  maintains the one-dimensional shear flow presented in the previous section, in the presence of  a vertical hydrostatic equilibrium density gradient imposed. The equilibrium is set so that the pressure gradient balances an external gravity force ($a_y$=$-g$), i.e., $-{dp}/{dy}-\rho g=0$. The analytic solution for the test case is,

\begin{equation}\label{IC_shear_hydrostatic}
\begin{aligned}
    u(x,y)&=U\sin(my) \\ 
    v(x,y)&=0 \\
    p(x,y)&=c_s^2\rho(x,y) \\
    \rho(x,y)&=\rho_0 \exp{(-gy/c_s^2)} \, .
\end{aligned}
\end{equation}

We run the simulation using these as initial conditions for $\sim 50$ sound crossing times, on a grid consisting of $64\times64$ zones. The simulation has dimensions $L_x=L_y=1$, $c_s=1$, $\rho_0=1$, and  $g=1$. The boundary conditions are $\rho(x,y=0)=\rho_0$, and $\rho(x,y=1)=\rho_0/e$, and are strictly periodic in the $x$ direction. We show the results of the integration in Figure \ref{fig:hydrostatic}. It is evident that the updated scheme maintains the analytic solution in the presence of the nonzero dissipation. 

\section{Numerical Assessment of Vorticity Preservation for the Full Euler Equations}
\label{sec:assessment}

\begin{figure}
\begin{center}
\includegraphics[angle=0,trim={.1cm .1cm .1cm .1cm},scale=.59]{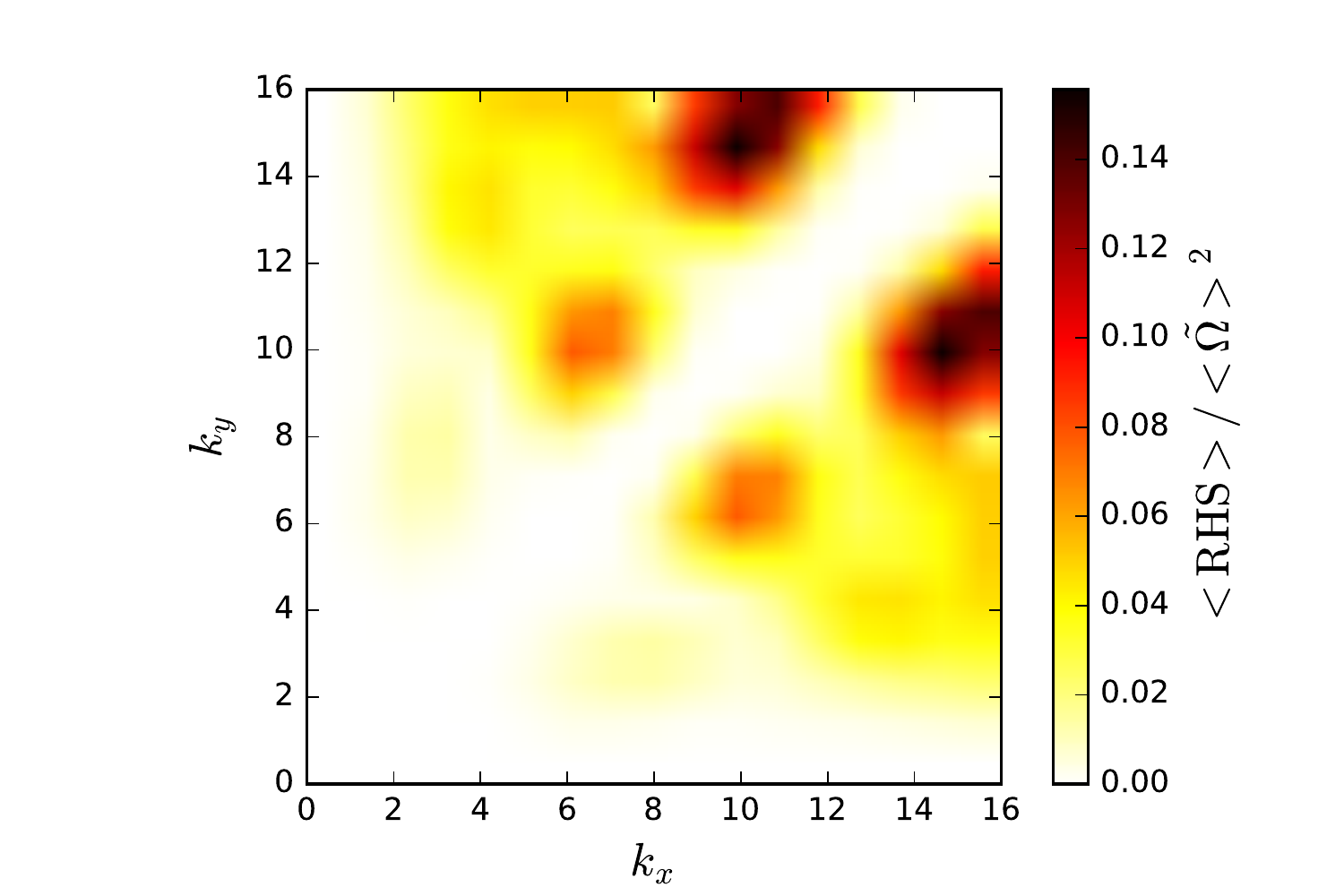}
\includegraphics[angle=0,trim={.1cm .1cm .1cm .1cm},scale=.59]{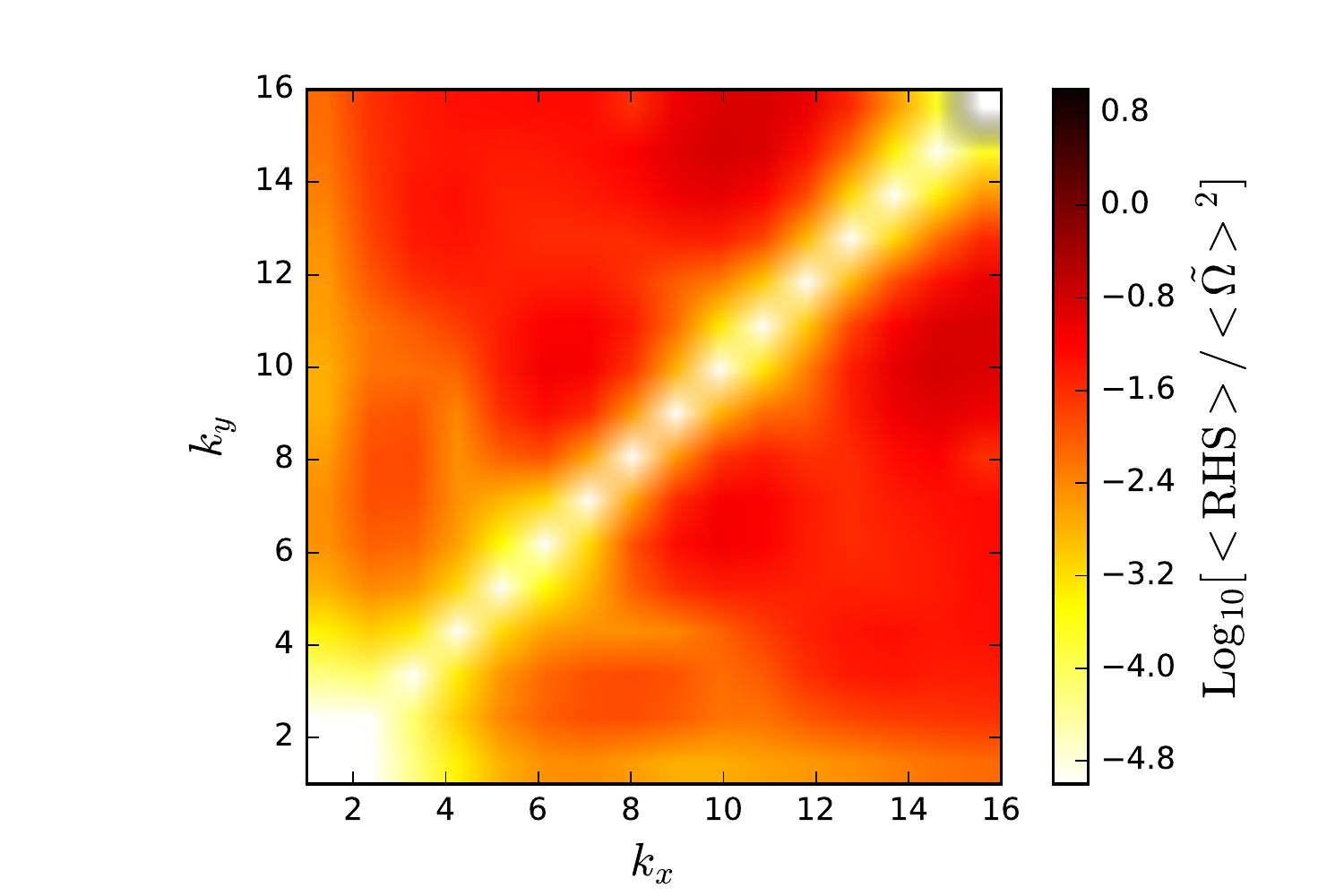}
\caption{A numerical assessment of vorticity preservation for discrete wavenumber, vortical modes up to the Nyquist frequency. We present the RMS of the residual (RHS) of the vorticity equation normalized by the RMS of the original seeded vorticity for a range of vortical modes. The top and bottom panels show this quantity on a linear and log scale respectively. The $x$ and $y$ axes correspond to the wavenumbers $k_x$ and $k_y$ of the seeded vortical mode. The $k_x,k_y=0$ modes correspond to unskewed shear layers, which have been shown analytically to be vorticity preserving. }
\end{center}
\label{fig:RMSRHS}
\end{figure}
In this section, we numerically quantify the extent to which the RBV2 scheme preserves vorticity for the full Euler equations. We do this by examining the dissipation in the vorticity evolution equation for a spectrum of vortical modes up to the Nyquist frequency. For the full Euler equations, given by Equation \ref{fulleuler}, the scheme \ref{LeratScheme} is vorticity preserving if (given by Definition 2 of \cite{Lerat2007})

\begin{equation}
    D_0\mu_0\tilde{\Omega}+D_1[D_1\mu(\rho uv)+D_2\mu(\rho v^2)]-D_2[D_1\mu(\rho u^2)+D_2\mu(\rho uv)]=0\, ,
\end{equation}
where now $\tilde{\Omega}=D_1(\rho v)-D_2(\rho u)$. Therefore the vorticity equation is given by Equation \ref{vorticiy_evolution}, and the RHS of which  is $\mathrm{RHS} =h_1D_1\widetilde{\mathrm{curl}}_1+h_2D_2\widetilde{\mathrm{curl}}_2$. Henceforth, we write
\begin{equation}
\begin{aligned}
    \tilde{q}^{(1)}&=[D_0\mu(\rho)+D_1(\rho u)+D_2( \rho v)]\\
    \tilde{q}^{(2)}&=[D_0\mu(\rho u)+D_1(\rho u^2+p)+D_2 (\rho u v)]\\
    \tilde{q}^{(3)}&=[D_0\mu(\rho v)+D_1(\rho uv)+D_2 (\rho v^2 +p)]\, .
\end{aligned}
\end{equation}

In this formalism, $\widetilde{\mathrm{curl}}_1 = D_1(\Phi_1 \tilde{q})^{(3)}-D_2(\Phi_1 \tilde{q})^{(2)}$.Therefore $ (\Phi_1 \tilde{q})^{(3)}=uv\tilde{q}^{(1)}+v\tilde{q}^{(2)}+u\tilde{q}^{(3)}$ and $ (\Phi_1 \tilde{q})^{(2)}=(u^2+c_s^2)\tilde{q}^{(1)}+2u\tilde{q}^{(2)}$. Similarly, $\widetilde{\mathrm{curl}}_2 = D_1(\Phi_2 \tilde{q})^{(3)}-D_2(\Phi_2 \tilde{q})^{(2)}$. To construct this,  $ (\Phi_2 \tilde{q})^{(3)}=(v^2+c_s^2)\tilde{q}^{(1)}+2v\tilde{q}^{(3)}$ and 
$ (\Phi_2 \tilde{q})^{(2)}=uv\tilde{q}^{(1)}+v\tilde{q}^{(2)}+u\tilde{q}^{(3)}$. Therefore

\begin{equation}
\begin{split}
    \widetilde{\mathrm{curl}}_1=D_1[uv\tilde{q}^{(1)}+v\tilde{q}^{(2)}+u\tilde{q}^{(3)}]
    -D_2[(u^2+c_s^2)\tilde{q}^{(1)}+2u\tilde{q}^{(2)}]\, ,
\end{split}
\end{equation}

and

\begin{equation}
\begin{split}
    \widetilde{\mathrm{curl}}_2=D_1[(v^2+c_s^2)\tilde{q}^{(1)}+2v\tilde{q}^{(3)}] 
    -D_2[uv\tilde{q}^{(1)}+v\tilde{q}^{(2)}+u\tilde{q}^{(3)}] \,.
\end{split}
\end{equation}

Then, for a uniform grid where $h_1=h_2=h$ the RHS simplifies to 

\begin{equation}\label{residuals}
\begin{split}
    \mathrm{RHS}=h[D_1D_1-D_2D_2][uv\tilde{q}^{(1)}+v\tilde{q}^{(2)}+u\tilde{q}^{(3)}]\\+hD_1D_2[(v^2-u^2)\tilde{q}^{(1)}-2u\tilde{q}^{(2)}+2v\tilde{q}^{(3)}]\, ,
\end{split}
\end{equation}
with $\mu_0=\mu^2$, $\mu=\mu_1\mu_2$, $D_0=\frac{1}{\Delta t}\bar{\Delta}$, $D_1=\frac{1}{\delta x}\delta_1\mu_2$ and $D_2=\frac{1}{\delta y}\delta_2\mu_1$, as defined in Section 2. If the dissipation in the RBV2 scheme exactly preservers vorticity, Equation \ref{residuals} should yield identically zero on the grid. Any deviations would signify that the scheme was dissipating vorticity. 

We calculate the residuals of the $\mathrm{RHS}$ from Equation \ref{residuals}. The results of this are shown in Figure \ref{fig:RMSRHS}. Using a $32\times32$ zone grid, we seed vortical perturbations of the form

\begin{equation}\label{RMS_ICS}
\begin{aligned}
    u(x,y)&=A\frac{k_x}{\sqrt{k_x^2+k_y^2}} \Re \{\exp{\Big(i\frac{2\pi}{L}(k_xx-k_yy)\Big)}\}\\ 
    v(x,y)&=A\frac{k_y}{\sqrt{k_x^2+k_y^2}}\Re\{ \exp{\Big(i\frac{2\pi}{L}(k_xx-k_yy)\Big)}\} \\
    p(x,y)&=c_s^2\rho(x,y) \\
    \rho(x,y)&=\rho_0  \, ,
\end{aligned}
\end{equation}
for a set of all integers $(k_x,k_y)\in \mathbb{Z}\times \mathbb{Z}$ from 1 to the nyquist wavenumber $N/2=16$. We evaluate the RHS of the discrete vorticity equation, and show the  root mean square of this quantity normalized by the initial vorticity, $<\mathrm{RHS}>/<\tilde{\Omega}>^2$ in Figure \ref{fig:RMSRHS}. We show the linear and log scale of this assessment, and in the linear scaling include the case where $k_x=k_y=0$, which correspond to the unskewed shear layers that we have analytically shown to be zero. It appears that RBV2 perfectly evolves vorticity for vortical modes of equal wavenumbers, e.g. $k_x=k_y$. The scheme dissipates vorticity when the wavenumbers are antisymmetric, e.g.$k_x\ne k_y$ , and the dissipation increases for higher wavenumbers. 

\section{Dynamical Interaction of Vortices in a Protoplanetary Disk}

\begin{figure}
\includegraphics[angle=0,trim={0.0cm 0.0cm 0.0cm 0.0cm},scale=.3]{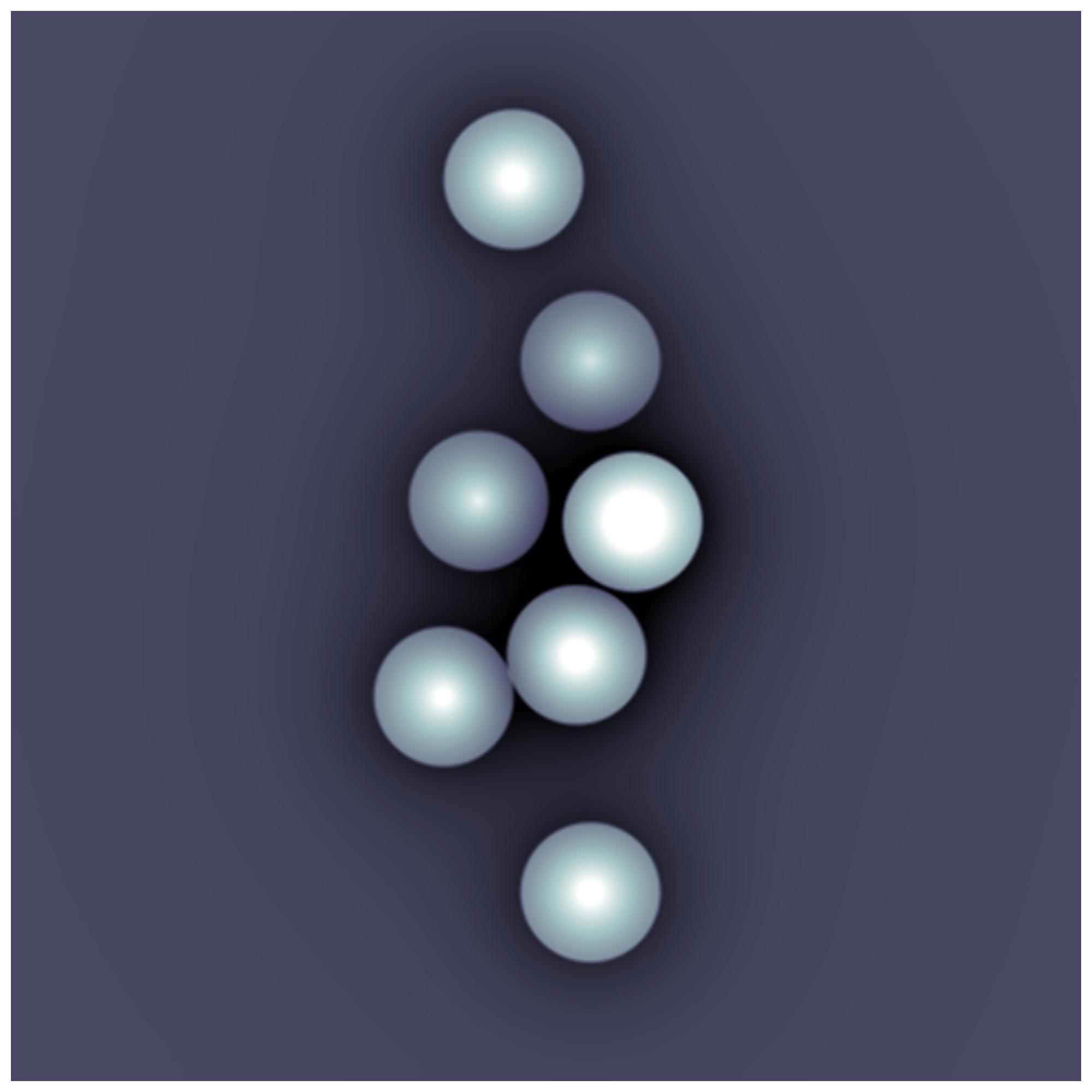}
\includegraphics[angle=0,trim={0.0cm 0.0cm 0.0cm 0.0cm},scale=.3]{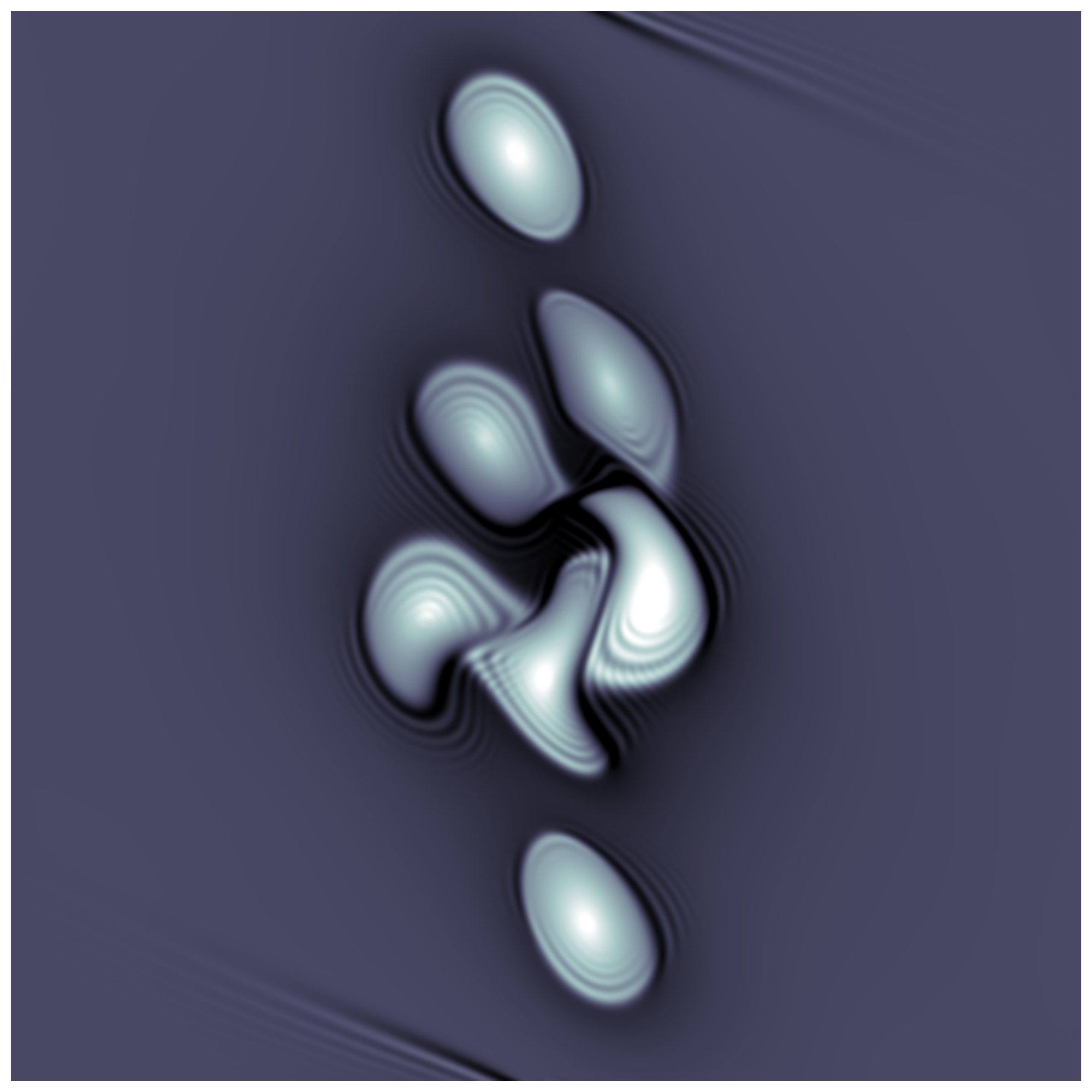}
\includegraphics[angle=0,trim={0.0cm 0.0cm 0.0cm 0.0cm},scale=.3]{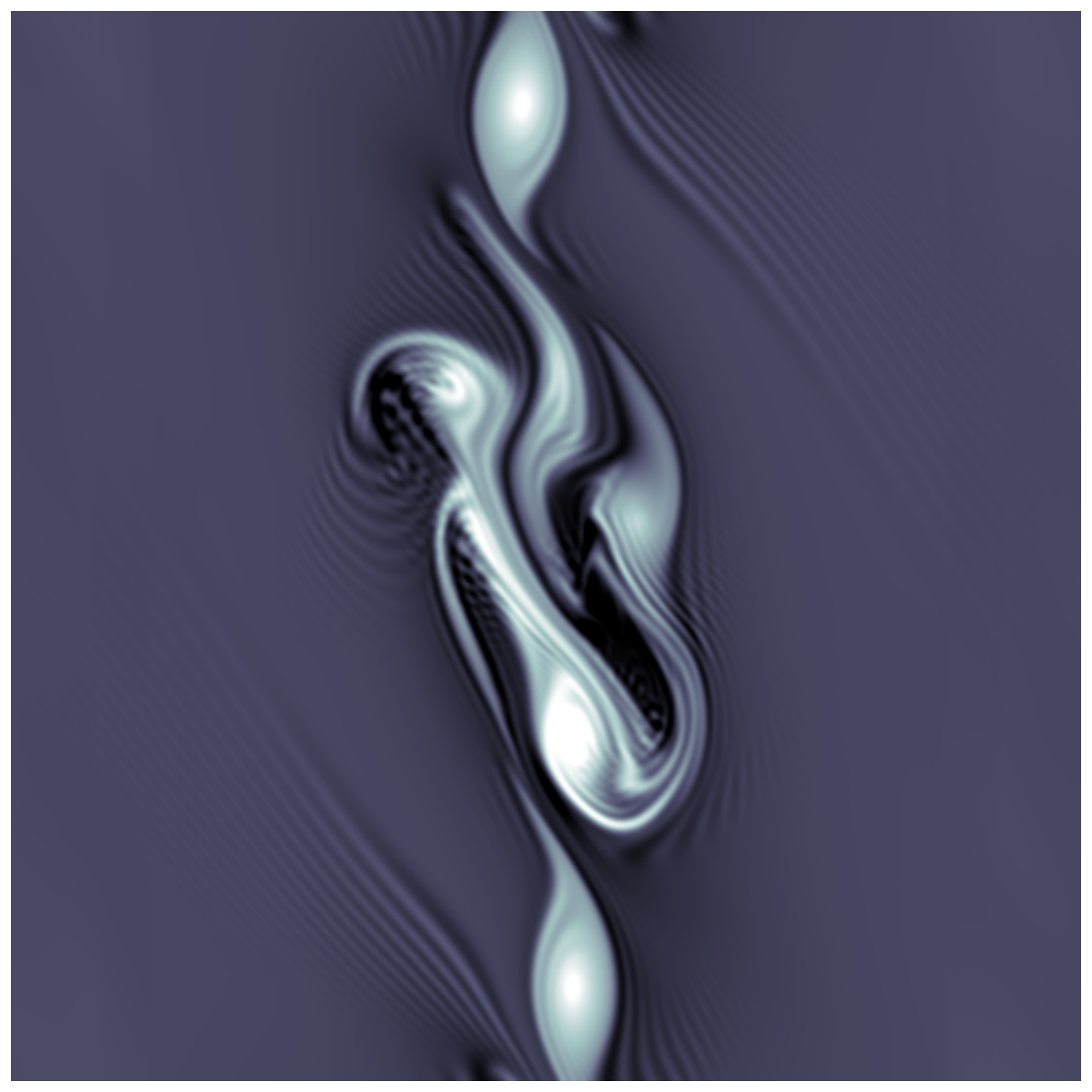}
\includegraphics[angle=0,trim={0.0cm 0.0cm 0.0cm 0.0cm},scale=.3]{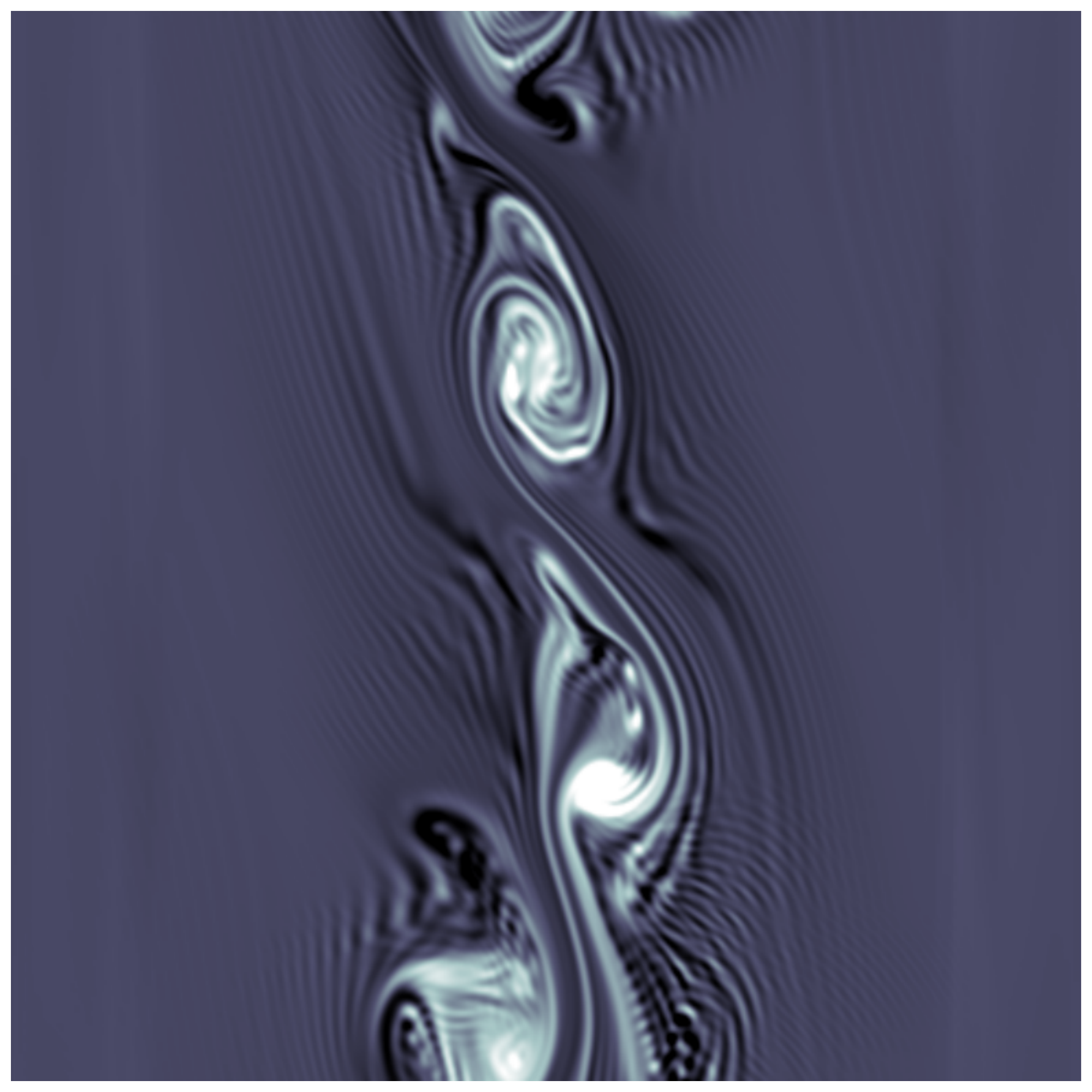}
\includegraphics[angle=0,trim={0.0cm 0.0cm 0.0cm 0.0cm},scale=.3]{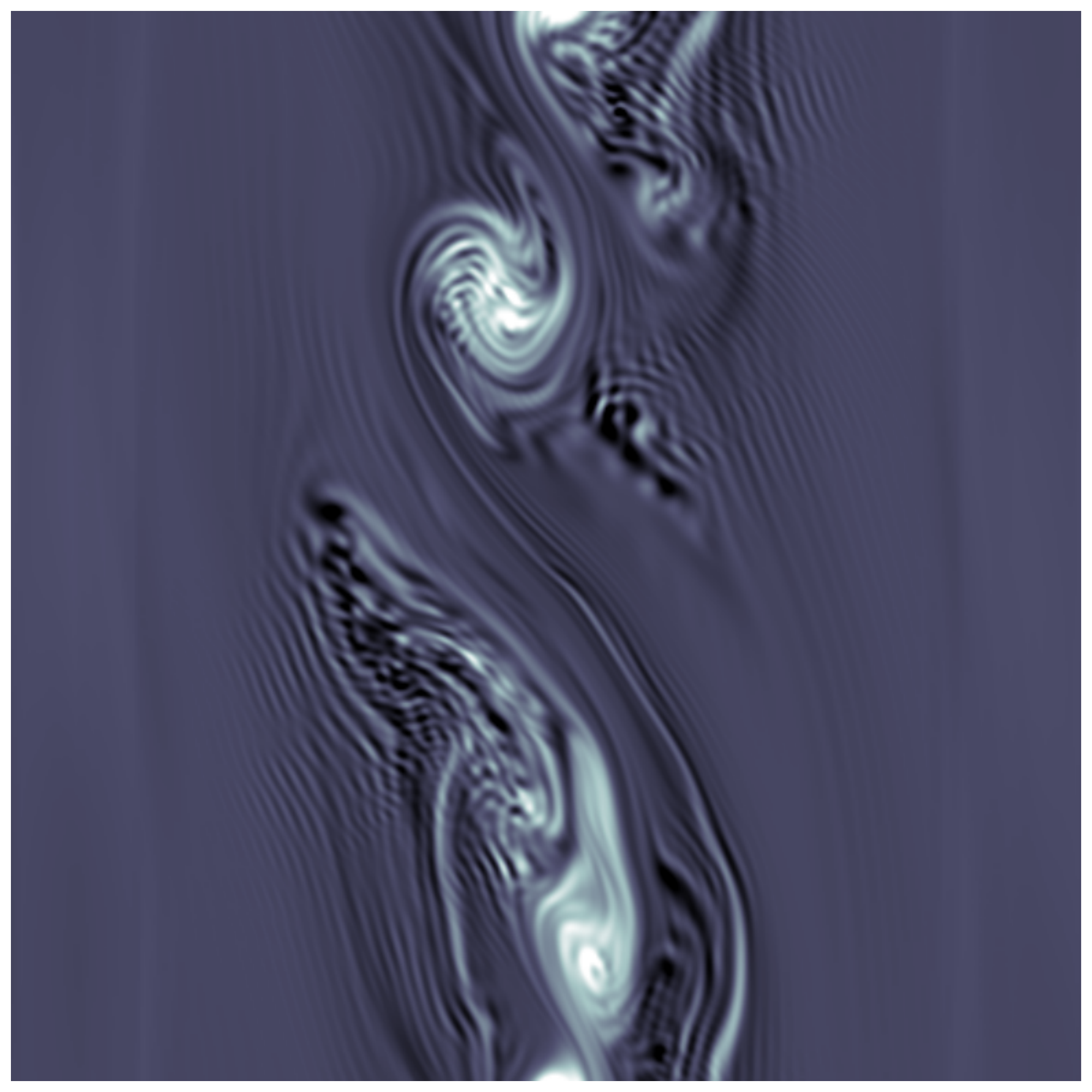}
\includegraphics[angle=0,trim={0.0cm 0.0cm 0.0cm 0.0cm},scale=.3]{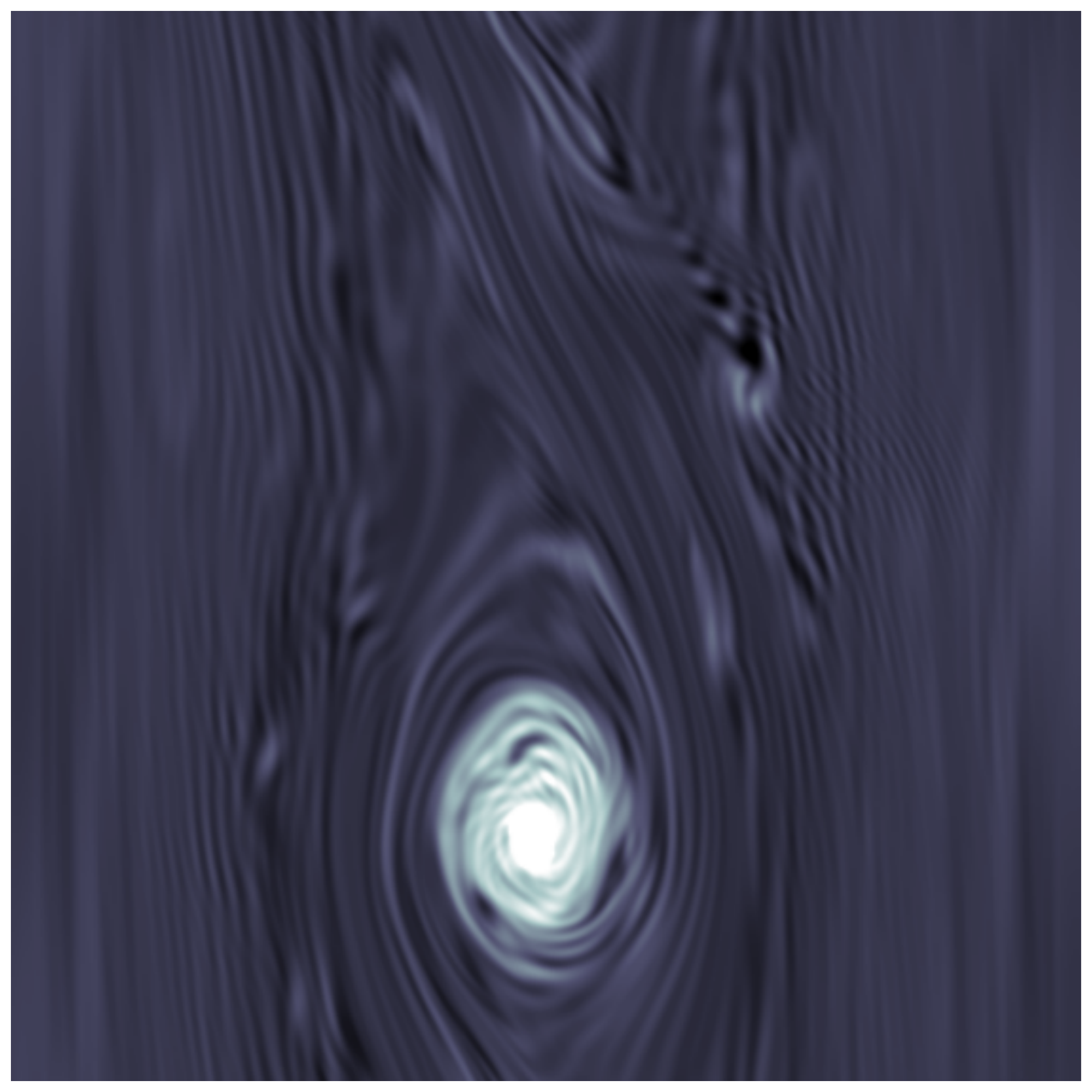}
\caption{The dynamical interaction of vortices in a protoplanetary disk. The color-scale corresponds to the vorticity in the shearing sheet. The simulation was run on 512$\times$512 zones, at a fiducial radius of $1$AU, with a sound speed $c_s=0.05$ and radial and azimuthal extent of $.04$AU. The Keplerian shear initially dominates the dynamics for the outer two vortices, and the five central vortices merge to form two new ones almost instantaneously. The three long-lived, stable vortices shear past each other for significant evolution, dynamically interacting and exchanging vorticity, before eventually coalescing into a single, much larger whirlpool.}
\label{fig:VortexMergers}
\end{figure}
There now exists mounting theoretical and observational evidence that vortices play  an important role in the local and global dynamics of protoplanetary disks. Therefore, developing computational algorithms that can accurately evolve flows replete with vortices and turbulence is critical to any theoretical understanding of them.  Vortices have been invoked as the sites for  planet formation due to the rapid settling of dust grains into their central regions \citep{Barge1995}. This process enhances the local surface density, and may trigger giant planet formation via gravitational instability \citep{Adams1995}.\citet{vandermarel2013} reported ALMA observations of the disk around the star  Oph IRS 48 that traced millimeter sized grains, and were consistent with a massive, anticyclonic dust trapping vortex. \citet{Cazzoletti2018} presented similar ALMA observations as evidence for a massive ($\ge0.3M_{Jup}$) vortex in the HD 135344B disk. Moreover, they argued that  the  vortex may be massive enough to launch spiral density waves, and could therefore be responsible for  the observed, \textit{global}, spiral structure.  \cite{abramowic1992} suggested that vortices in hot accretion disks may explain the variability in their X-ray temporal power spectra. Similar temporal variability is observed in young stellar objects \citep{Flaherty2016}, which may also be caused by vortices. 

From a theoretical standpoint, vortices and hydrodynamical turbulence in protostellar disks may be important drivers of  angular momentum transport. In the ideal MHD limit, the magnetorotational instability \citep{Balbus1991,Balbus1991b} produces both the  magnitude and outward directionality  of the angular momentum transport necessary for the viscous evolution of  gas in an accretion disk. However, protoplanetary disks are so weakly ionized that magnetic fields may not affect the motion of the neutral medium, so there remains interest in searching for a purely hydrodynamical mechanism for the outward transport of angular momentum.  \citet{Barranco2005} discovered that vortices that form naturally above the midplane from disk perturbations  are robust for hundreds of orbit, and could produce significant  angular momentum transport. Self-gravitating, cooling instabilites \citep{Gammie2001}, baroclinic instabilities \citep{Klahr2003} and zombie vortex instabilities \citep{Marcus2015} may also transport angular momentum (for a recent review, see \cite{Lyra2018}). 
 
Here, we present a simulation of the dynamical interaction of seven vortices in a protoplanetary disk. This provides an example of how the RBV2 algorithm, updated to handle external body forces, will be useful for studying turbulent astrophysical flows. We perform a simulation in the Cartesian shearing box formalism, which is employed to examine local phenomenon at a scale which captures the shear, yet omits the global curvilinear geometry, described in detail in \citet{Hawley1995}. For a Keplerian disk, the hydrodynamical equations of motions expanded around a fiducial disk radius are

\begin{equation}\label{eq:SheerSheetEOM}
\frac{\partial {\bf v}}{\partial t} + {\bf v}\cdot\vec{\nabla} {\bf v}=-\frac{\vec{\nabla} P}{\rho}-2\boldsymbol{\Omega}\times {\bf v} +3|\boldsymbol{\Omega}|^2 x \hat{x} \, ,
\end{equation}
where ${\bf v}$, $P$, $\rho$ and $\boldsymbol{\Omega}$ are the local velocity, pressure, density and rotational velocity. In addition to the descripton of the implentation in Section 4.1 in \citet{Seligman2017}, the operator $\tilde{{\boldsymbol q}}$ has been updated at each grid cell to include the appropriate coriolis term, $-2\boldsymbol{\Omega}\times {\bf v}$, and   tidal expansion  of the effective centrifugal and gravitational potential from the central object term, $3|\boldsymbol{\Omega}|^2 x \hat{x}$. The seeded vortices have the functional form proposed in \cite{Adams1995}, and used in Section 4.2 and 4.3 of \cite{Seligman2017}. The simulation  was run on 512$\times$512 zones, at a fiducial radius of $1$AU, with a sound speed $c_s=0.05$ and radial and azimuthal extent of $.04$AU. 

Figure \ref{fig:VortexMergers} shows the dynamical interaction of  seven anticyclonic vortices in a protoplanetary disk as rendered with the updated RBV2 algorithm. Only anticyclones are expected to be long-lived in disks, because they have the same sense of rotation as that of the local shear, which is \textit{opposite} that of the global rotation. The same shear which enforces the rotation will rip apart  cyclonic vortices \citep{kida1981,Barranco2005}. In the absence of the background shear and the additional forces present in the rotating frame, vortices  that have the same spin orientation and are close enough together will eventually merge \citep{Cerretelli2003}. As demonstrated in the second and third panels of Figure \ref{fig:VortexMergers}, the five central vortices dynamically interact very rapidly. The Keplerian shear  dominates the dynamics of the two furthest vortices, driving the upper left vortex upwards and lower right vortex downwards. In the fourth panel, the five central vortices have coalesced into two new vortices with substantially more sub-structure. Due to the periodic boundary condition in the azimuthal direction, the two furthest vortices have also begun merging at this stage. The three long-lived vortices continue to shear past each other for $\sim 10$ orbits, dynamically interacting, exchanging vorticity and shedding Rossby waves that dramatically effect the sub-structure in the surrounding flow. After $\sim 18$ sound crossing times, the three vortices coalesce into a single, much larger whirlpool.

\section{Conclusion}

\citet{Lerat2007} presented a novel approach to the accurate evolution of vorticity by insisting on vorticity preservation in idealized hydrodynamical cases on a multi-dimensional grid. In this paper, we investigate the vorticity-preserving properties of the resultant RBV2 algorithm. We evaluate how the scheme handles several idealized hydrodynamical tests of the full Euler equations. We propose an adjustment to the dissipation that retains the vorticity-preserving qualities, while accurately incorporating external body forces, and demonstrate that the adjusted scheme sustains a hydrostatic equilibrium. 
We present a novel numerical assessment of vorticity dissipation for discrete wave-number, vortical modes. We find that for the full Euler equations, RBV2 perfectly preserves vorticty for modes with symmetric wavenumbers, and that the error increases with asymmetry. We then perfrom shearing sheet simulations of vortex interactions in a protoplanetary disk, to demonstrate the utility of the updated algorithm for modeling astrophysical fluid flows. 

The RBV2 algorithm has proven to exhibit remarkably efficient and accurate results for test problems involving turbulence and vortices. Given its minimal dissipation of vorticity, we plan use  the  algorithm to simulate astrophysical environments  where turbulence is expected to dominate the dynamics. One very promising example of this is a protoplanetary disk, where turbulence and vortices may dictate the accretion process \citep{Barranco2005}. With the advent of the proposed update to include body forces, we believe that the  scheme will prove extremely useful for elucidating the nature of and interaction between vortices in Jupiter's polar region, where the small Rossby number \citep{Barranco2005} implies that the Coriolis deflection dominates the dynamics.
\acknowledgments

This material is based upon work supported by the National
Aeronautics and Space Administration through the NASA
Astrobiology Institute under Cooperative Agreement Notice
NNH13ZDA017C issued through the Science Mission Directorate.
We acknowledge support from the NASA Astrobiology
Institute through a cooperative agreement between NASA
Ames Research Center and Yale University.

We thank Professor Alain Lerat for extensive correspondence regarding the contents of this paper. We thank Professor Gregory Laughlin for extensive support and insightful conversations. 

\software{https://github.com/DSeligman/MAELSTROM2D}

\bibliography{refs}



\end{document}